\documentclass[showpacs,amsmath,amssymb,aps,showkeys,floatfix,
prd,a4paper,twocolumn]{revtex4-1} 
\usepackage[utf8]{inputenc} %
\usepackage{verbatim} 
\usepackage[dvips]{graphicx}
\usepackage{dcolumn}
\usepackage{bm}
\usepackage{epsfig}
\usepackage{amsfonts}
\usepackage{amssymb,amscd}
\usepackage{color} 
\usepackage{lineno}
\usepackage[normalem]{ulem}

\newcommand{\be}{\begin{equation}}
\newcommand{\ee}{\end{equation}}
\newcommand{\ben}{\begin{eqnarray}}
\newcommand{\een}{\end{eqnarray}}

\begin{document}

\title{$\Upsilon$ suppression  in a hadron gas}
\author{ L. M. Abreu\footnote{luciano.abreu@ufba.br}}   
\affiliation{
Instituto de F\'isica, Universidade Federal da Bahia, 
Campus Universit\'ario de Ondina, 40170-115, Bahia, Brazil
}
\author{ F.~S.~Navarra\footnote{navarra@if.usp.br}}
 \affiliation{
Instituto de  F\'{i}sica, Universidade de S\~ao Paulo, Rua do Mat\~ao 1371,
05508-090 São Paulo, SP, Brazil
}
\author{ M.~Nielsen\footnote{mnielsen@if.usp.br} }
 \affiliation{
Instituto de  F\'{i}sica, Universidade de S\~ao Paulo, Rua do Mat\~ao 1371,
05508-090 São Paulo, SP, Brazil\\
SLAC Nacional Acelerator Laboratory, Stanford University, Stanford, 
California 94309, USA}

\begin{abstract}

{In this work we study the interactions of bottom mesons which lead 
to $\Upsilon$ production and absorption in hot hadronic matter.  
We use effective Lagrangians to calculate the $\Upsilon$ production cross section 
in processes such as $ \bar{B}^{(*)} + B^{(*)}  \to \Upsilon + (\pi, \rho)$  and
also the $\Upsilon$ absorption cross section in the corresponding inverse
processes. We update and extend previous calculations by Lin and Ko, introducing 
anomalous interactions. The obtained cross sections are used as input to solve 
the rate equation which allows us to follow the time evolution of the $\Upsilon$ 
multiplicity. In contrast to previous conjectures, our results suggest that 
the interactions in the hadron gas phase strongly reduce  
the $\Upsilon$ abundance. }

\end{abstract}

\maketitle

\section{Introduction}

One of the most interesting predictions of QCD is that  strongly interacting 
matter  undergoes a phase transition to a deconfined state at sufficiently 
high temperatures. The medium composed of quarks and gluons in 
this deconfined state is referred to as the quark-gluon plasma (QGP) and it 
has been observed in heavy ion collisions at RHIC \cite{qgpdisc} and at LHC
\cite{rev-qgp}. 

Heavy quark bound states are believed to be reliable probes of 
the quark gluon plasma. 
In the QGP, once the heavy quarkonium states are formed they are expected  
to unbind due to the strong interactions with partons in the medium through   
a QCD Debye  screening mechanism \cite{matsatz,sup-rev-rapp}.  
Above a certain temperature, the more weakly 
bound states,  such as $\Upsilon(3S)$, are expected to unbind more completely 
compared to the more strongly bound states, e.g. $\Upsilon(1S)$. At even higher  
temperatures, more of the weakly bound states are expected to dissolve. In   
the experiment, this sequential unbinding (also referred to as melting) of  
quarkonium states is expected to be observed as a sequential suppression of  
their yields. The  suppression of heavy quarkonium states was accordingly 
proposed as the 
smoking-gun signature of the phase transition, and its sequential pattern 
as a probe of the medium temperature \cite{hf-rev}. 

In the early days most of the attention was devoted to the suppression of
charmonium states  in collider experiments at SPS and RHIC. However, even 
after decades of intense efforts, the experimental  
observations are not yet completely understood. The suppression of 
$\psi (1S)$ does 
not increase from SPS to RHIC, or from RHIC to LHC, even though in each   
change of accelerator 
the center-of-mass energy is increased by one order of magnitude. The most  
accepted explanation for this ``unsuppression'' is 
that  heavy quarks, evolving independently in the QGP,  recombine forming  bound 
states. This process is called recombination or regeneration 
\cite{hf-rev,regraf}.   
It is supposed to take place in the hot plasma and hence to affect mostly the  
charmonium states produced with  transverse momenta typical of the quark-gluon  
fluid.  Indeed, the relative (compared to the scaled  
pp baseline) reduction of the $J/\psi$ 
multiplicity  measured in AA collisions at low $p_T$ 
is significantly smaller at LHC energies than at RHIC energies. This is  
consistent with the regeneration mechanism since the larger  charm          
production cross section at LHC enhances the probability of recombination.  
The situation changes at high $p_T$, 
where the suppression rises as the collision energy increases, revealing 
that the  $J/\psi$ yield is less sensitive to 
recombination~\cite{Abelev:2013ila,Adam:2016rdg,Zha:2017xsm}.

While charmonium states have been extensively studied as  QGP probes,    
bottomonium states were not explored so much, even though the
$ b \bar{b}$  family of states provides experimentally more robust and   
theoretically cleaner probes. Moreover, bottomonium states are regarded   
as better probes because recombination effects are believed to be much less   
significant than  
in the charmonium  case.  Although the recombination effect is expected 
to increase  
for bottomonia from RHIC to LHC energies, it is predicted to remain small 
\cite{durapphe,regebot,chineses,emerapp,mike1,mike2}. 

From the experimental side, the CMS detector has excellent capabilities for 
muon detection and provides measurements of the $\Upsilon$ family which 
enable the  accurate analysis of bottomonium \cite{leonardo} production.   
For this reason,  the main interest may be shifted to the suppression of 
bottomonium states at  LHC energies.
The first indication of $\Upsilon$ suppression in heavy ion collisions was 
reported by CMS in 2011 \cite{cms}. Later it was also observed by the STAR 
Collaboration at RHIC \cite{upsup-star}. The $\Upsilon(2S)$  and $\Upsilon(3S)$  
resonances in PbPb collisions were seen to be more strongly suppressed than the 
$\Upsilon(1S)$ (compared with the pp result), showing the expected sequential 
suppression pattern \cite{leonardo}.

The  most recent data on prompt $J/\psi$ \cite{cms-psi}  and 
$\Upsilon$ \cite{leonardo,cms-ups-18,cms-ups-17,cms-ups-17a}             
suppression in the most central Pb Pb collisions at small rapidities and 
small $p_T$, show that:
\be
R_{AA} (J/\psi) \simeq 0.28 \pm 0.03
\label{raapsi}
\ee
and
\be
R_{AA} (\Upsilon(1S)) \simeq 0.38 \pm 0.05
\label{raaups}
\ee
These factors  are very weakly dependent on the collision energy 
$\sqrt{s}_{NN}$. Although they are close to each other,  they may 
be the result of a quite different dynamics. 

After the QGP cooling and hadronization there is a hadron gas (HG) phase. 
Apart from being a reasonable assumption,  the 
existence of this phase seems to be necessary to correctly reproduce 
\cite{hg-pbm} the multiplicities of $K^*$ and $\rho$  measured by the ALICE 
Collaboration \cite{hg-alice,hg-kst,hg-rho}. Heavy quarkonium is produced at 
the beginning of the heavy ion collision. Then 
it may be destroyed and regenerated both in the quark gluon plasma and in the 
subsequent hadron gas. The observed $\Upsilon$ suppression has been explained 
mostly with models which take into account only what happens during the QGP 
phase. In this work we address the contribution of the hadron gas phase to the 
$\Upsilon$ production and absorption.

In the literature,  there is a large number of works on  
quarkonium  interactions with light mesons in a hot hadron gas using different 
approaches
(for a short and recent compilation of references on charmonium interactions, 
see \cite{akhematonani}). 
Many of these works investigate the $J/\psi$-light meson reactions based on   
effective hadron Lagrangians~\cite{akhematonani,psipi-ft1,psipi-oh,
kmnn}.  After a long series of works, different groups found a similar  
value of the $J/\psi-\pi$ cross section, which is  
close to the value obtained with QCD sum rules \cite{psipi-sr}. 
In \cite{akhematonani}, we have used all the known charmonium-light hadron  
absorption cross sections (together with the inverse  interactions in which 
charmonium is produced)  as input to solve the rate equation which governs 
the time evolution of $J/\psi$ abundance in a hadron gas. The  effective  
Lagrangian approach will  be employed also to the bottomonium in the next 
sections. 

In contrast to the $J/\psi$ case,  the number of studies about  
the $\Upsilon$ interactions with light hadrons is much smaller. 
In fact, to the best of our knowledge, the paper by Lin and Ko, 
Ref.~\cite{Upsilon1}, is the only one to give an estimate        
of  the cross sections for scattering of $\Upsilon$ by pions and 
$\rho$ mesons in a hot hadron gas. In that work 
the authors used a  hadronic Lagrangian based on the SU(5) flavor   
symmetry. Including form factors  with a cutoff parameter of 1 or 2 
GeV at the interaction vertices, they found that the values of      
$\sigma_{\pi \Upsilon}$  and $\sigma_{\rho \Upsilon}$   are about 8 mb  
and 1 mb, respectively.  However, due to the large kinematic threshold, 
their thermal averages at a temperature of 150 MeV are                
both only about 0.2 mb. They then conclude speculating that  the   
absorption of directly produced $\Upsilon$  by comoving hadrons is 
unlikely to be important. 

In view of the recent  theoretical and experimental progress on    
$\Upsilon$ physics, we believe that it is time to update and extend the 
calculation of Ref.~\cite{Upsilon1}. In the present work we will 
contribute to this subject extending the analysis                 
performed in Ref.~~\cite{akhematonani} to the bottomonia sector:  
we investigate the interactions of $\Upsilon$ with the surrounding 
hadronic medium composed of the    
lightest pseudoscalar meson ($\pi$) and the lightest vector meson ($\rho$). We  
calculate the cross sections for processes such as $ \bar{B}^{(*)} + B^{(*)}  
\to \Upsilon + (\pi, \rho)$  scattering and their inverses, within the 
effective hadron Lagrangian  framework. We improve the 
previous calculation introducing anomalous interactions.
The obtained cross sections are used as input to solve the       
rate equation which allows us to follow the time evolution of the $\Upsilon$ 
multiplicity.

The importance of the anomalous vertices has been earlier mentioned in  
different contexts. 
For example, in Ref. \cite{psipi-oh} the $J/\psi$ absorption cross   
sections by pions and $\rho$ mesons  
were evaluated for several processes producing $D$ and $D^*$ mesons   
in the final state. 
The authors found that the $J/\psi \pi \to  D^*  \bar{D}$ cross section   
obtained with the 
exchange of a $D^*$  meson in the t-channel, which involves   the   
anomalous $D^* D^* \pi$  
coupling, was around 80 times bigger than the one obtained with a  
$D$ meson exchange in the t-channel. In Ref. \cite{rocaoset} the    
authors studied the radiative decay modes 
of the $f_0(980)$ and $a_0(980)$ resonances, finding that the diagrams   
involving anomalous 
couplings were quite important for most of the decays. More recently,   
in Refs. \cite{nosx1,nosx2}
it was shown that the inclusion of anomalous interactions produces 
significant changes in the $X(3872) \pi$ cross section. 

This work is organized as follows. In Section ~\ref{CrSec} we present an 
overview of the effective Lagrangian formalism and calculate 
the cross section for $\Upsilon$ production and absorption. The results 
obtained for the  
thermally averaged cross sections are exhibited  and discussed  in 
Section~\ref{AvCrSec}. After that, Section~\ref{TimeEv} is 
dedicated to the analysis of $\Upsilon$ abundance in heavy ion               
collisions. Finally, in Section~\ref{Conclusions} we summarize the 
results and conclusions.

\section{ Interactions between $\Upsilon $ and light mesons}
\label{CrSec}

\subsection{Effective Lagrangian Formalism}

In the present study the reactions involving the $\Upsilon$ production and 
absorption will be analyzed in the effective 
field theory approach. Accordingly, 
we follow Refs.~\cite{akhematonani,psipi-ft1,psipi-oh} and employ 
 the couplings between light- and heavy-meson fields within the framework of 
an $SU(4)$ effective formalism, in which the 
vector mesons are identified as the gauge 
bosons, and the interaction Lagrangians are given by
\begin{eqnarray}
\mathcal{L}_{PPV} & = & -ig_{PPV}\langle V^\mu[P,\partial_\mu P]\rangle ,  
\nonumber \\
\mathcal{L}_{VVV} & = & i g_{VVV} \langle \partial_\mu V_\nu 
\left[ V^{\mu}, V^{\nu} 
\right] \rangle , 
\nonumber \\
\mathcal{L}_{PPVV} & = & g_{PPVV}\langle P V^\mu[V_\mu , P]\rangle ,  
\nonumber \\
\mathcal{L}_{VVVV} & = & g_{VVVV}\langle V^\mu V^\nu [V_\mu , V_\nu]\rangle , 
\label{Lagr1}
\end{eqnarray}
where the indices $PPV$ and $VVV$, $PPVV$ and $VVVV$ denote the type of vertex 
incorporating pseudoscalar ($P$) and vector ($V$) meson fields in the couplings
~\cite{psipi-ft1,psipi-oh}  and                     
$g_{PPV}$, $g_{VVV}$, $g_{PPVV}$ and $g_{VVVV}$ are the respective 
coupling constants. The symbol $\langle \ldots \rangle$ stands for the 
trace over $SU(4)$-matrices. 
$V_\mu$ represents a $SU(4)$ matrix, which is parametrized by 16 vector-meson 
fields including the 15-plet and singlet 
of $SU(4)$, 
\begin{eqnarray}
V_\mu = \begin{pmatrix}	 
\frac{\omega}{\sqrt{2}}+\frac{\rho^0}{\sqrt{2}} & \rho^+ & K^{*+} & 
\bar B^{*0} \\
	\rho^{-} & \frac{\omega}{\sqrt{2}}-\frac{\rho^0}{\sqrt{2}} & 
K^{*0} & B^{*-} \\
	K^{*-} & \bar K^{*0} & \phi & B^{*-}_s \\
	B^{*0} & B^{*+} & B^{*+}_s & \Upsilon 
\end{pmatrix}_\mu ;
\label{eq:2}
\end{eqnarray}
$P$ is a matrix containing the 15-plet of the pseudoscalar meson fields, 
written in the physical basis in which $\eta$, 
$\eta ^{\prime}$ mixing is taken into account,
\begin{eqnarray}
 P = \begin{pmatrix}
	\frac{\eta}{\sqrt{3}}+\frac{\eta^{\prime}}{\sqrt{6}}+\frac{\pi^0}{\sqrt{2}} & \pi^{+} & K^{+} & \bar B^{0} \\
	\pi^{-} & \frac{\eta}{\sqrt{3}}+\frac{\eta^{\prime}}{\sqrt{6}}-\frac{\pi^0}{\sqrt{2}} & K^{0} & B^{-} \\
	K^{*-} & \bar K^{*0} & -\frac{\eta}{\sqrt{3}}+\sqrt{\frac{2}{3}}\eta^{\prime} & B^{-}_s \\
	B^{0} & B^{+} & B^{+}_s & \eta_b 
\end{pmatrix} . \nonumber 
\end{eqnarray}

We also consider anomalous parity interactions in addition to the couplings 
given above. 
The anomalous parity interactions can be 
described in terms of the gauged Wess-Zumino action~\cite{psipi-oh}, and  
are written as
\begin{eqnarray} 
\mathcal{L}_{PVV} & = & - g_{PVV} \varepsilon^{\mu\nu\alpha\beta} \langle 
\partial_\mu V_\nu \partial_\alpha V_\beta P \rangle ,  \nonumber \\
\mathcal{L}_{PPPV} & = & - i g_{PPPV} \varepsilon^{\mu\nu\alpha\beta} 
\langle V_\mu (\partial_{\nu} P) (\partial_{\alpha} P) (\partial_{\beta} P) \rangle , 
\nonumber \\
\mathcal{L}_{PVVV} & = & i g_{PVVV} \varepsilon^{\mu\nu\alpha\beta} \left[  
\langle V_\mu  V_\nu  V_\alpha \partial_{\beta} P \rangle  \right. 
\nonumber \\
& & \left. +  \frac{1}{3} \langle V_\mu (\partial_\nu V_\alpha)  V_\beta P \rangle 
\right].
\label{Lagr2}
\end{eqnarray}
The $g_{PVV}$, $g_{PPPV}$, $g_{PVVV}$ are the coupling constants of 
the $PVV$, $PPPV$ and $PVVV$ vertices, 
respectively~\cite{akhematonani,psipi-ft1,psipi-oh}. 

The effective Lagrangians given in Eqs.~(\ref{Lagr1}) 
and~(\ref{Lagr2}) allow us to study the following $\varphi \Upsilon $ 
absorption  processes
\begin{eqnarray}
(1) \;\; \varphi  \Upsilon & \rightarrow &  \bar{B} B, \nonumber \\
(2) \;\; \varphi  \Upsilon & \rightarrow &   \bar{B} ^{\ast} B  , 
\nonumber \\
(3) \;\; \varphi \Upsilon &  \rightarrow &  \bar{B} ^{\ast} B ^{\ast} , 
\label{proc1}
\end{eqnarray}
where in the initial states $\varphi$ stand for pions and $\rho$ mesons. 
The process $\varphi \Upsilon  
\rightarrow    \bar{B} B ^{\ast} $ has the same
cross section  as the process (2) in Eq.~(\ref{proc1}).
In the present approach, the diagrams considered to compute the 
amplitudes of the 
processes above are of two types: one-meson exchange and contact graphs.  
They are shown in Fig.~\ref{Diag1} and~\ref{Diag2} for those with pions 
and $\rho$, respectively. 
\begin{widetext}
  
\begin{figure}[!ht]
    \centering
       \includegraphics[width=10.5cm]{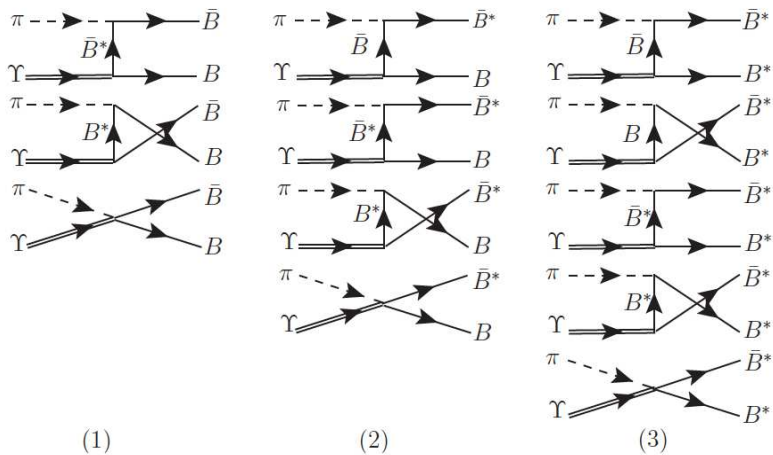}
        \caption{Diagrams contributing to the processes:   
(1) $ \pi  \Upsilon \rightarrow  \bar{B} B $, (2) 
$\pi  \Upsilon \rightarrow  \bar{B}^{\ast} B  $, and (3) 
$\pi  \Upsilon \rightarrow  \bar{B}^{\ast} B ^{\ast}$.  }
\label{Diag1}
\end{figure}

\begin{figure}[!ht]
    \centering
       \includegraphics[width=10.5cm]{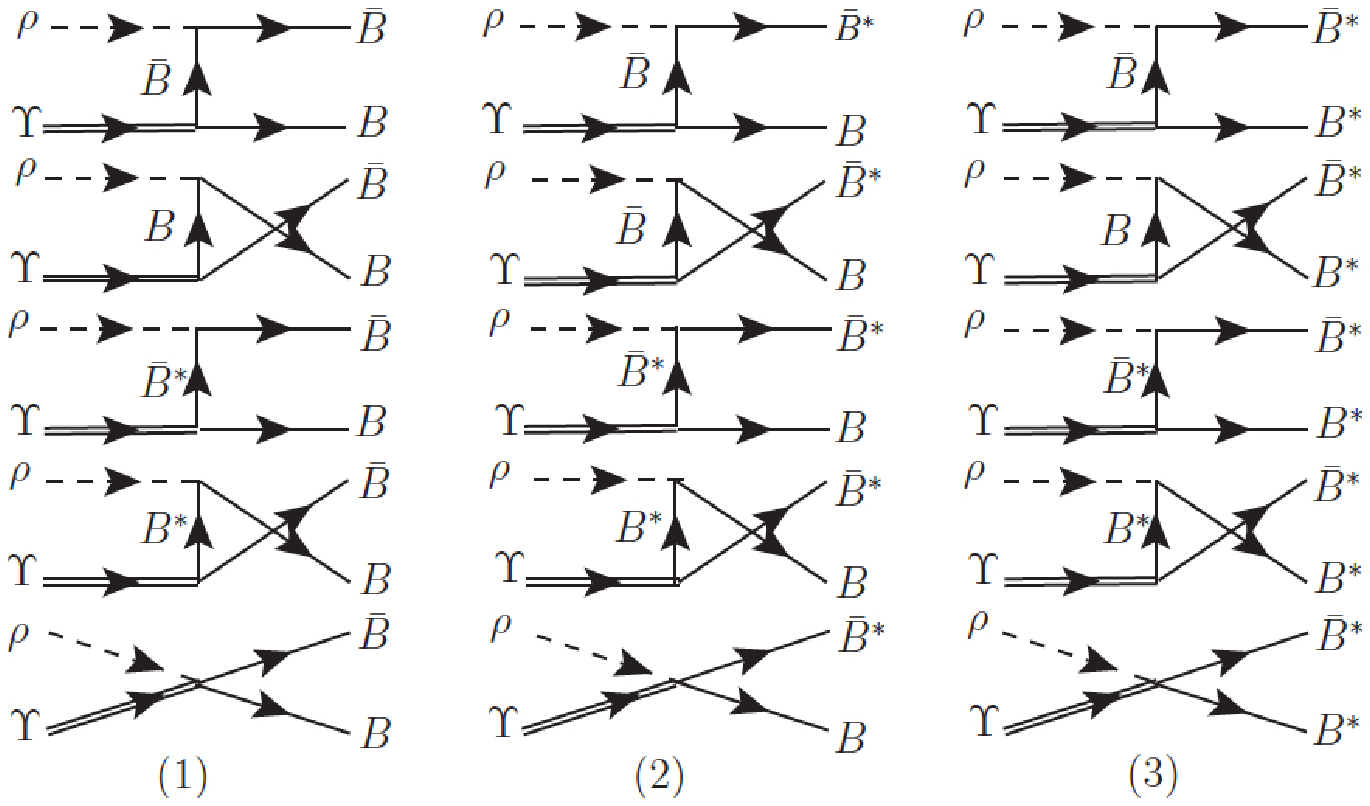}     
        \caption{Diagrams contributing to the processes: 
(1) $ \rho  \Upsilon \rightarrow  \bar{B} B $, (2)    
$\rho  \Upsilon \rightarrow  \bar{B}^{\ast} B  $, and 
(3) $\rho  \Upsilon \rightarrow  \bar{B}^{\ast} B ^{\ast}$. }
\label{Diag2}
\end{figure}

\end{widetext}

We define the invariant amplitudes for the processes (1)-(3) in 
Eq. (\ref{proc1}) 
involving $\varphi = \pi$ meson as 
 \begin{eqnarray}
   \mathcal{M}_1 ^{(\pi)} & = &  \sum _{i} \mathcal{M}_{1i}^{(\pi)\mu} 
\epsilon_{ \mu} (p_1), \nonumber \\
   \mathcal{M}_2 ^{(\pi)} & = &  \sum _{i} \mathcal{M}_{2i}^{(\pi)\mu \nu} 
\epsilon_{\mu} (p_1) \epsilon_{\nu} ^{\ast} (p_3) . \nonumber \\
   \mathcal{M}_3 ^{(\pi)} & = &  \sum _{i} 
\mathcal{M}_{3i}^{(\pi)\mu \nu \lambda} \epsilon_{\mu} (p_1) 
\epsilon_{\nu} ^{\ast} (p_3) \epsilon_{\lambda} ^{\ast} (p_4), 
      \label{amplpi}
 \end{eqnarray}
while for the ones involving $\varphi = \rho$ meson we have
\begin{eqnarray}
   \mathcal{M}_1 ^{(\rho)} & = &  \sum _{i} \mathcal{M}_{1i}^{(\rho)\mu \nu} 
\epsilon_{ \mu} (p_1) \epsilon_{ \nu} (p_2), \nonumber \\
   \mathcal{M}_2 ^{(\rho)} & = &  \sum _{i} \mathcal{M}_{2i}^{(\rho)\mu \nu \lambda} 
\epsilon_{\mu} (p_1) \epsilon_{\nu} (p_2)  \epsilon_{\lambda} ^{\ast} (p_3) . \nonumber \\
   \mathcal{M}_3 ^{(\rho)} & = &  \sum _{i} 
\mathcal{M}_{3i}^{(\rho)\mu \nu \lambda \delta} \epsilon_{\mu} (p_1) \epsilon_{\nu} (p_2) 
\epsilon_{\lambda} ^{\ast} (p_3) \epsilon_{\delta} ^{\ast} (p_4).
      \label{amplrho}
 \end{eqnarray}
 In the above equations, the sum over $i$ represents the sum over all diagrams 
contributing to the respective amplitude; $p_j$  
denotes the momentum of particle $j$, with particles 1 and 2 standing for initial 
state mesons, and particles 3 and 4 for final 
state mesons; $\epsilon_{ \mu} (p_j)$ is the polarization vector related to the 
respective vector particle $j$.   The specific expressions of amplitudes 
$\mathcal{M}_i ^{(\pi)}  $ and $\mathcal{M}_i ^{(\rho)}  $ in the present case  are analogous to the ones 
given in Ref.~\cite{psipi-oh} involving $\varphi J/ \psi  \rightarrow  D ^{(\ast)} \bar{D} ^{(\ast)} $. So, 
we will  not reproduce here the explicit expressions of the invariant 
amplitudes.  These expressions can be found in  Ref.~\cite{psipi-oh}, taking into account the replacement of masses and coupling 
constants labeled with charmed mesons and $J/\psi$ by similar quantities labeled with bottomed mesons and $\Upsilon$, respectively.

The isospin-spin-averaged cross section for the processes in Eq. (\ref{proc1}) is 
defined in the center of mass (CM) frame as
\begin{eqnarray}
  \sigma_r ^{\left(\varphi \right)}(s) 
= \frac{1}{64 \pi^2 s }  \frac{|\vec{p}_{f}|}{|\vec{p}_i|}  \int d \Omega 
\overline{\sum_{S, I}} 
|\mathcal{M}_r  ^{\left(\varphi \right)} (s,\theta)|^2 ,
\label{eq:CrossSection}
\end{eqnarray}
where $r = 1,2,3$ labels  $\varphi -  \Upsilon $ absorption processes according to 
Eqs.~(\ref{amplpi}) and~(\ref{amplrho}); $\sqrt{s}$ is the CM energy;  $|\vec{p}_{i}|$ and $|\vec{p}_{f}|$ 
denote the three-momenta of initial and final particles in the CM frame, respectively; 
the symbol $\overline{\sum_{S,I}}$ represents the sum over the spins and isospins of 
the particles in the initial and final state, weighted by the 
isospin and spin degeneracy 
factors of the two particles forming the initial state for the reaction $r$, i.e. 
\begin{eqnarray}
\overline{\sum_{S,I}}|\mathcal{M}_r|^2 & = & \frac{1}{g_1 g_2 } 
\sum_{S,I}|\mathcal{M}_r|^2, \label{eq:DegeneracyFactors}
\end{eqnarray}
with $g_1 = (2I_{1i,r}+1)(2S_{1i,r}+1), g_2 = (2I_{2i,r}+1)(2S_{2i,r}+1) $ being the 
degeneracy factors of the initial particles 1 and 2. 

We have employed in the computations of the present work the isospin-averaged 
masses:  $m_{\pi} = 137.3 $ MeV, $m_{\rho} = 775.2 $ MeV, $m_B = 5279.4 $ MeV, 
$m_{B^{\ast}} = 5324.7$ MeV, $m_{\Upsilon} = 9460.3 $ MeV.  
The values of coupling constants appearing in the expressions of the 
amplitudes $\mathcal{M} ^{(\pi)}$ and $\mathcal{M} ^{(\rho)}$ are given in 
Table~\ref{couplconst} ~\cite{psipi-oh,Upsilon1}. 

\begin{table}
\caption{Values of coupling constants~\cite{psipi-oh,Upsilon1}. } 
\begin{center}
\begin{tabular}{c|c}
\hline
\hline
Coupling constant  & Value \\
\hline
\hline
$g_{B^{\ast} B \pi}$ &  24.9 \\
$g_{\Upsilon B B}$ &  13.3 \\
$g_{\Upsilon B^{\ast} B^{\ast}}$ & 13.3 \\
$g_{ B B \rho}$ &  2.52 \\
$g_{ B^{\ast} B^{\ast} \rho}$ & 2.52 \\
$g_{ \Upsilon B^{\ast} B \pi}$ & 165.6 \\
$g_{ \Upsilon B B \rho}$ &  67.03 \\
$g_{ \Upsilon B^{\ast} B^{\ast} \rho}$ & 33.5 \\
\hline
$g_{ B^{\ast} B^{\ast} \pi}$ & 9.39 GeV${}^{-1}$\\
$g_{ \Upsilon B^{\ast} B }$ &  2.51 GeV${}^{-1}$\\
$g_{ B^{\ast} B \rho}$ &  1.84 GeV${}^{-1}$\\
$g_{ \Upsilon B B \pi}$ & 44.8 GeV${}^{-3}$\\
$g_{ \Upsilon B^{\ast} B^{\ast} \pi}$ & 31.25 GeV${}^{-1}$\\
$h_{ \Upsilon B^{\ast} B^{\ast} \pi}$ & 31.25 GeV${}^{-1}$\\
$g_{ \Upsilon B^{\ast} B \rho}$ & 6.33 GeV${}^{-1}$\\
$h_{ \Upsilon B^{\ast} B \rho}$ & 6.33 GeV${}^{-1}$ \\
\hline
\hline
\end{tabular}
\end{center}
\label{couplconst}
\end{table}

We have also included form factors in the vertices when 
evaluating the cross sections, defined as \cite{Pearce:1990uj,Oh:2002vg,Ronchen:2012eg}: 
\begin{eqnarray}
	F_3 & = & \left( \frac{ n \Lambda^4}{ n \Lambda ^4 + (p^2 - m_{ex} ^2 ) ^2} \right)^n , \;\; \nonumber \\
	F_4 & = & \left( \frac{ n \Lambda^4}{ n \Lambda ^4 + \left[ (p_1 + p_2 ) ^2 - (m_3 + m_4)^2 \right] } \right)^n ,
  \label{formfactor}
\end{eqnarray}
where $F_3$ and $F_4$ are the form factor for the three-point and four-point 
vertices, respectively; $p$ is the fou-momentum of the exchanged particle of mass $m_{ex}$ for a vertex involving a $t$- or $u$-channel 
meson exchange; $m_3$ and $m_4$ are the final state meson masses. 
The cutoff $\Lambda$ and $n$ parameters are chosen to be $\Lambda = 5.0 $ GeV and $n \rightarrow \infty $ for all 
vertices, which gives Gaussian form factors with witdh $ 25.0$ GeV${}^{2}$ .

\subsection{$\Upsilon$ production and absorption cross sections}

\begin{figure}[th]
\centering
\includegraphics[width=8.0cm]{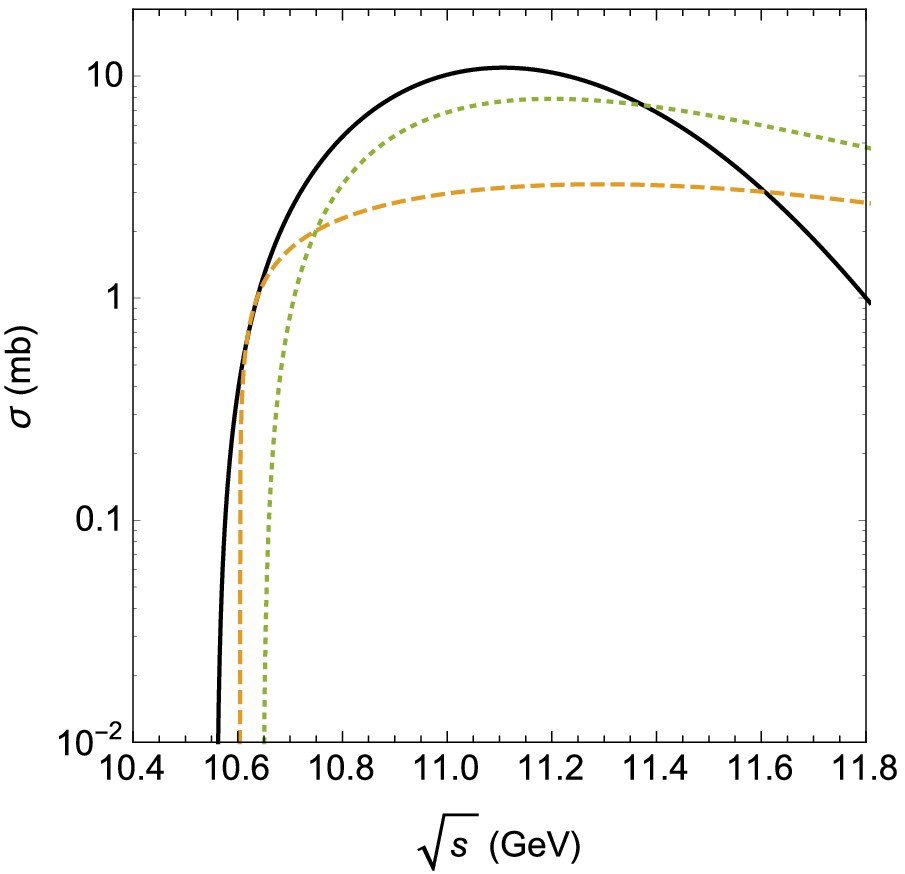}
\\
\includegraphics[width=8.0cm]{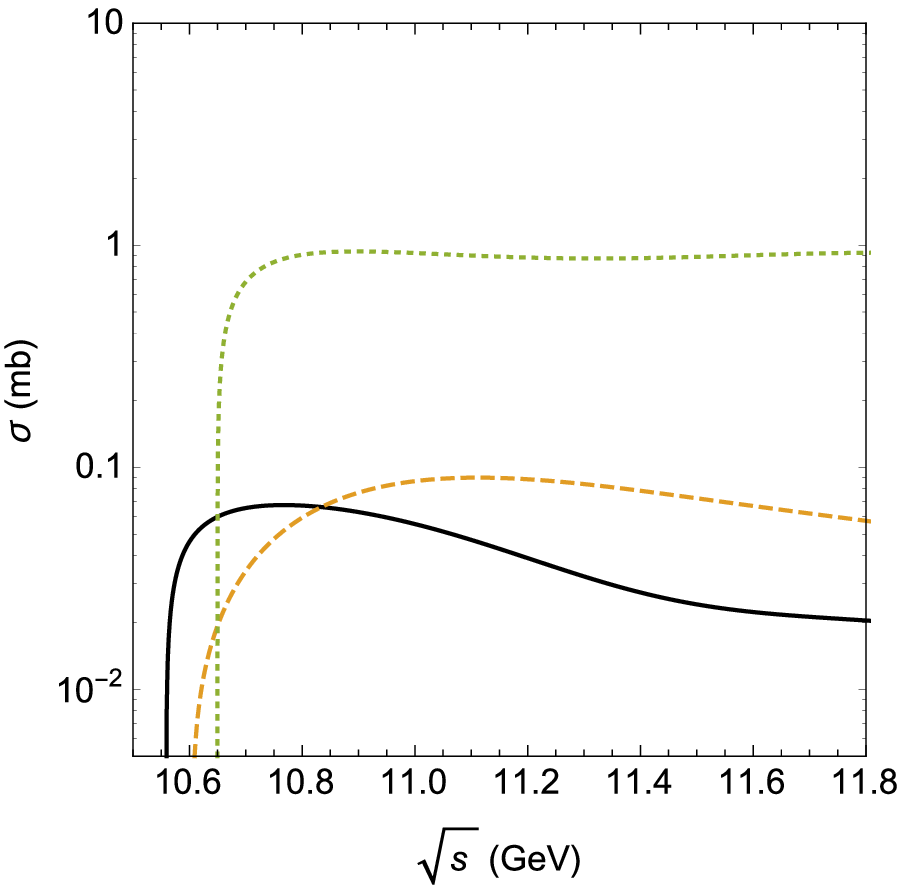}
\caption{$\Upsilon$ absorption cross sections in different processes 
as a function of the CM energy $\sqrt{s}$.  
Top panel:   $\pi  \Upsilon$  in the initial state.   
Bottom panel:  $\rho \Upsilon$  in the initial state. 
Solid, dashed and dotted lines represent the 
$\pi (\rho) \Upsilon  \rightarrow  \bar{B} B       $,  
$\pi (\rho) \Upsilon  \rightarrow  \bar{B}^{\ast} B        $ and 
$\pi (\rho) \Upsilon  \rightarrow  \bar{B}^{\ast} B^{\ast} $
reactions, respectively.  }
\label{CrSecUpsilon}
\end{figure}

On the top panel of Fig.~\ref{CrSecUpsilon} the $\pi \Upsilon $   
absorption cross sections for the $ \pi \Upsilon \rightarrow  
\bar{B} B, \bar{B}^{\ast} B  $ and  $\bar{B}^{\ast} B  ^{\ast}$
reactions  are plotted as a function of the CM energy $\sqrt{s}$.    
We see that the cross sections can be considered to be approximately   
of the same order of magnitude in the
 range $10.6 $ GeV $\le \sqrt{s} \le 11.8$ GeV, differing by about a   
factor 1.5-3. 
The magnitude of the reaction  $ \pi \Upsilon \rightarrow \bar{B}^{\ast} B$ 
is in agreement with previous 
calculations reported in Ref.~\cite{Upsilon1}, which is based in $SU(5)$   
symmetry, using different form factors, cutoffs and coupling constants and 
without anomalous terms. The authors of Ref.~\cite{Upsilon1} did not include  
some of the processes with final states  $\bar{B} B  $ 
and  $\bar{B}^{\ast} B  ^{\ast}$.

The cross sections of the processes  $ \rho \Upsilon  \rightarrow   
\bar{B} B, \bar{B}^{\ast} B  $ and  $\bar{B}^{\ast} B  ^{\ast}$  
are plotted as 
a function of $\sqrt{s}$ on the bottom panel of Fig.~\ref{CrSecUpsilon}.  
In this case the cross section for $ \rho \Upsilon  
\rightarrow \bar{B}^{\ast} B  ^{\ast}$ is larger than the others by about 
one order of magnitude.  As expected, the $\rho - \Upsilon $ reactions 
have smaller cross sections than those       
initiated by pions. The findings above are also in relative agreement 
with the previous calculations  reported in  
Ref.~\cite{Upsilon1}, although  only the processes which end with 
$ \bar{B} B$ 
and  $\bar{B}^{\ast} B  ^{\ast}$ have been  considered in ~\cite{Upsilon1}.  
Again, we believe that the differences are  
due to different choices in the form factors and cutoff values, and the 
absence of anomalous parity interactions.

	
\begin{figure}[th]
\centering
\includegraphics[width=8.0cm]{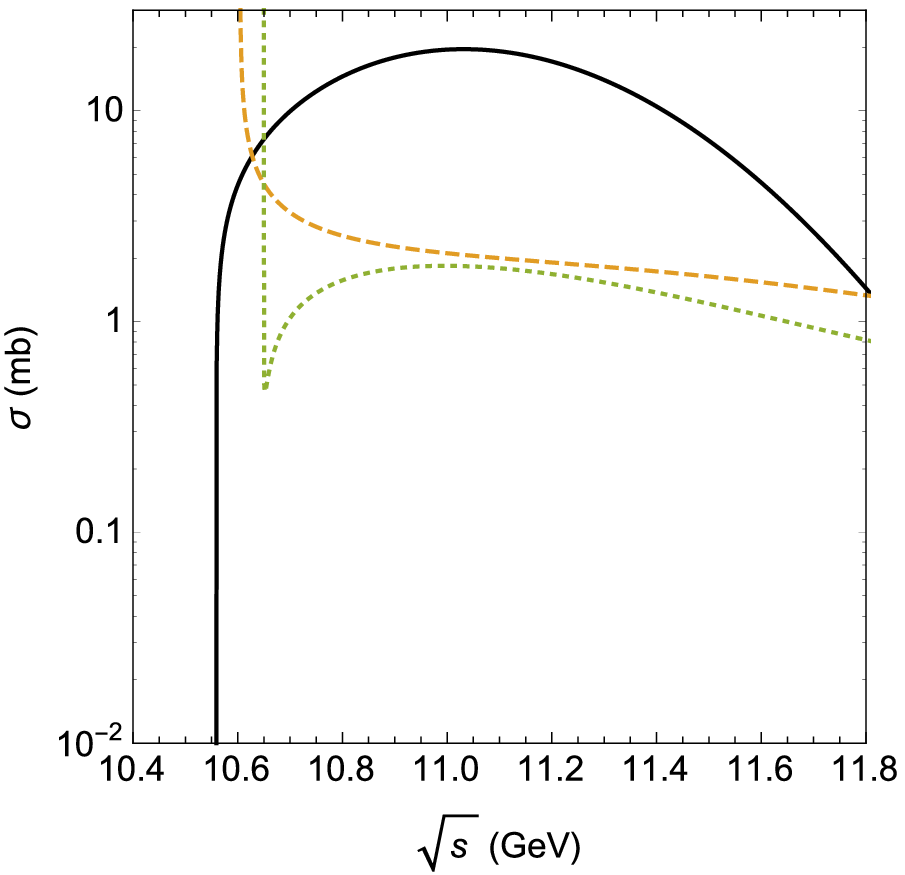}
\\
\includegraphics[width=8.0cm]{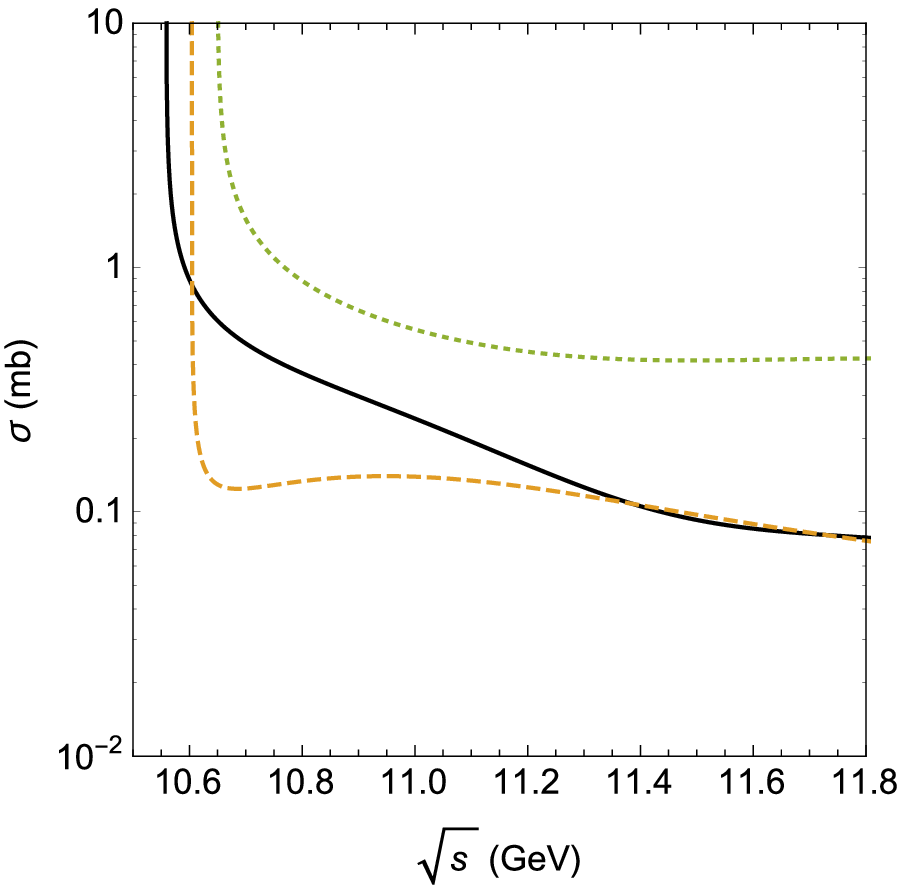}
\caption{$\Upsilon$ production cross sections in different processes 
as a function of the CM energy $\sqrt{s}$.  
Top panel:   $\pi  \Upsilon$  in the final state.   
Bottom panel:  $\rho \Upsilon$  in the final state. 
Solid, dashed and dotted lines represent the 
$  \bar{B} B   \rightarrow  \pi (\rho) \Upsilon   $,  
$ \bar{B} ^{\ast} B  \rightarrow \pi (\rho) \Upsilon      $ and 
$ \bar{B}^{\ast} B^{\ast}  \rightarrow \pi (\rho) \Upsilon $
reactions, respectively.  }
\label{CrSecUpsilonInv}
\end{figure}

For completeness, we now calculate the cross sections of the inverse 
processes, which can be 
obtained from the direct processes through the use of detailed balance 
(see for example Eq.~(48) of the last article of  Ref.~\cite{psipi-ft1}). 
In the top (bottom) panel of 
Fig.~\ref{CrSecUpsilonInv} the $ \pi (\rho) \Upsilon $ production cross  
sections  for the $  \bar{B} B   \rightarrow  \pi (\rho) \Upsilon  ,   
\bar{B} B^{\ast}  \rightarrow \pi (\rho) \Upsilon $ and 
$  \bar{B}^{\ast} B^{\ast}  \rightarrow \pi (\rho) \Upsilon $  
reactions are plotted as a function of the CM energy $\sqrt{s}$.

  
From these Figures  we can see that: i) processes 
which start or end with $\pi \Upsilon$ have larger cross sections. 
ii) Excluding the low energy region (which will be much less relevant for 
phenomenology), the $\Upsilon$ production and absorption cross sections are 
close to each other in almost all channels. 
Therefore, taking into account that the $\Upsilon$ absorption and production cross sections  have
comparable magnitudes, the computation of thermally averaged cross sections is an essential step to determine the final 
abundance of $\Upsilon$'s. This will be done in next Section.

\section{Thermally averaged cross sections}
\label{AvCrSec}
The thermally averaged cross section for a given process 
$a b \rightarrow c d$ is defined as~\cite{Koch,Cho:2010db,Cho:2015qca,nosx1}
\ben
\langle \sigma_{a b \rightarrow c d } v_{a b}\rangle &  = & 
\frac{ d^{3} \mathbf{p}_a  d^{3}
\mathbf{p}_b f_a(\mathbf{p}_a) f_b(\mathbf{p}_b) \sigma_{a b \rightarrow c d } 
\,\,v_{a b} }{ d^{3} \mathbf{p}_a  
d^{3} \mathbf{p}_b f_a(\mathbf{p}_a) f_b(\mathbf{p}_b) }
\nonumber \\
& = & \frac{1}{4 \alpha_a ^2 K_{2}(\alpha_a) \alpha_b ^2 K_{2}(\alpha_b) } 
\nonumber \\
& & \times \int _{z_0} ^{\infty } dz  K_{1}(z) \,\,\sigma (s=z^2 T^2) 
\nonumber \\
& & \times \left[ z^2 - (\alpha_a + \alpha_b)^2 \right]
\left[ z^2 - (\alpha_a - \alpha_b)^2 \right],
\nonumber \\
\label{thermavcs}
\een
where $v_{ab}$ represents the relative velocity of initial two interacting 
particles $a$ and $b$; the function $f_i(\mathbf{p}_i)$ is the Bose-Einstein   
distribution (of particles of species i), which depends on the temperature    
$T$; $\alpha _i = m_i / T$, $z_0 = max(\alpha_a + \alpha_b,\alpha_c 
+ \alpha_d)$, and $K_1$ and $K_2$ the modified Bessel functions of second kind.

\begin{figure}[th]  
\centering  
\includegraphics[width=8.0cm]{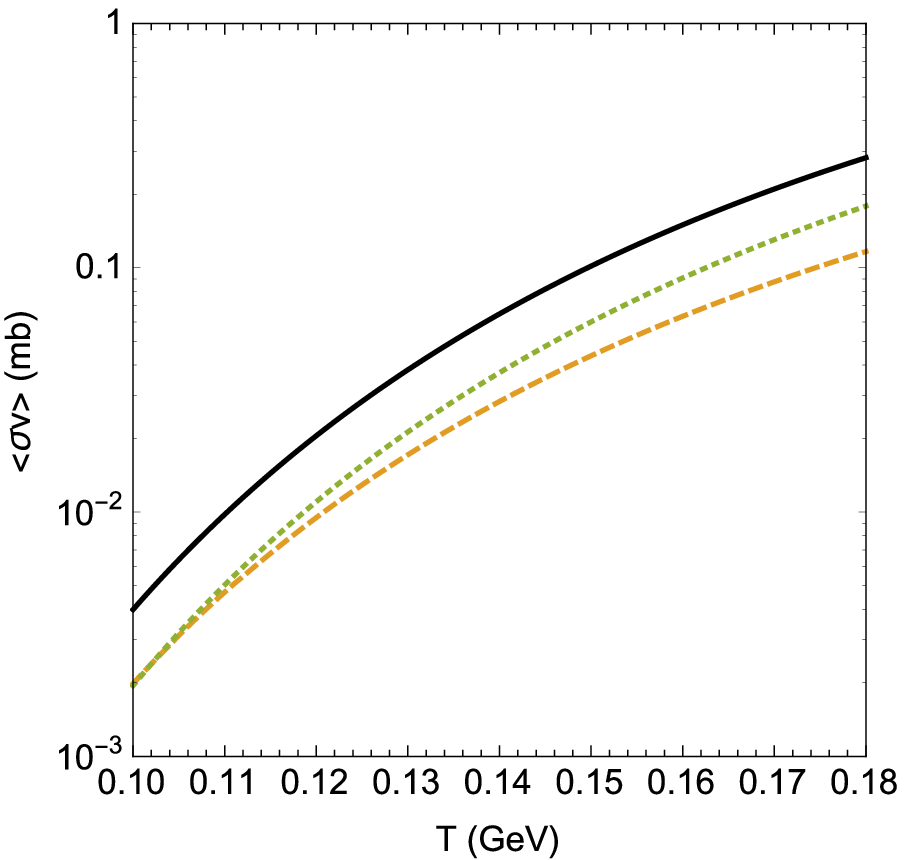}  
\\  
\includegraphics[width=8.0cm]{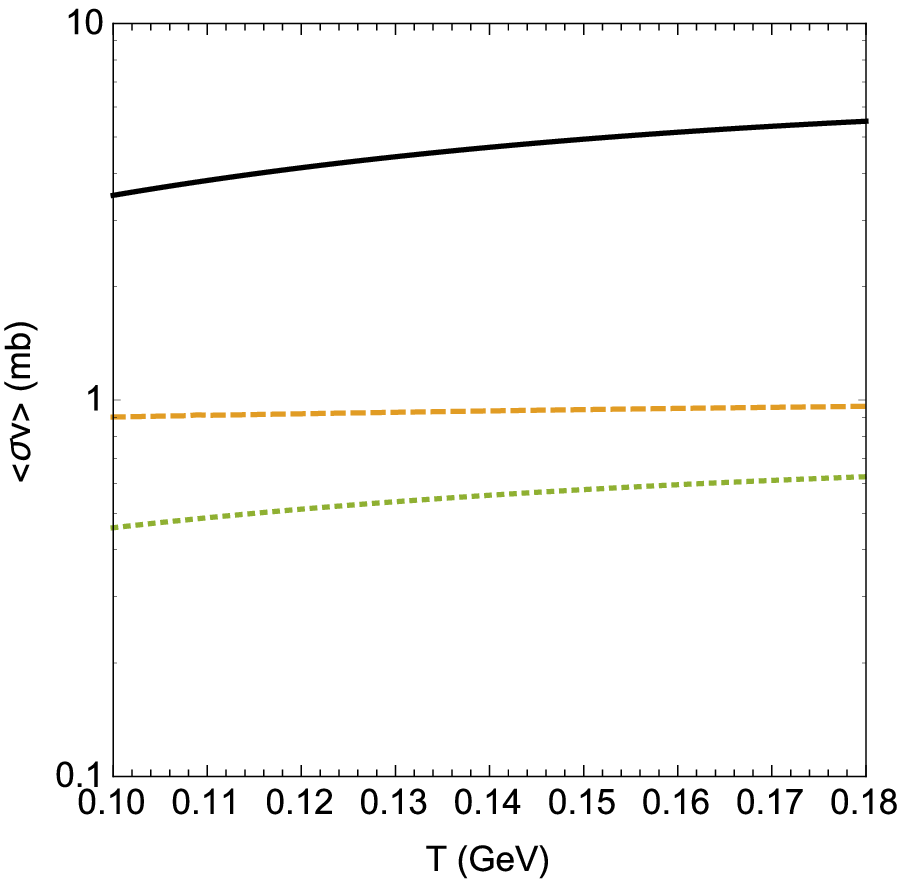}  
\caption{Thermally averaged cross sections for $\pi \Upsilon $ 
absorption and production as a function  of the temperature.  
Top panel:   $\pi  \Upsilon$  in the initial state.   
Bottom panel:  $\pi \Upsilon$  in the final state. 
Solid, dashed and dotted lines represent the reactions 
with $ \bar{B} B $,  $  \bar{B} ^{\ast} B $ and  
$  \bar{B}^{\ast} B^{\ast} $, respectively, in final or initial state.}
\label{thermalpi}  
\end{figure}

In Fig.~\ref{thermalpi} we plot the thermally 
averaged cross sections for  $\pi \Upsilon $ absorption (upper panel) and 
production (lower panel) via the 
processes discussed in previous section. We can see that reactions which 
start or end with $ \bar{B} B$ have greater magnitudes.  
But the main point here is that the production reactions have larger cross 
sections than the absorption ones. 

\begin{figure}[th]  
\centering  
\includegraphics[width=8.0cm]{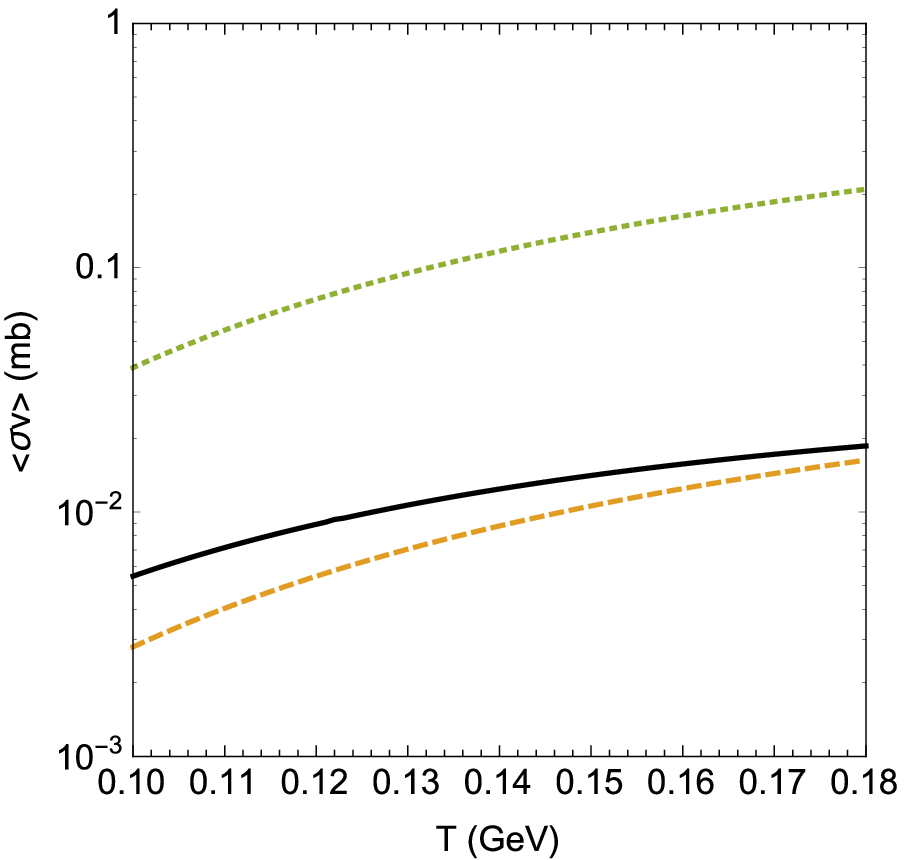}  
\\  
\includegraphics[width=8.0cm]{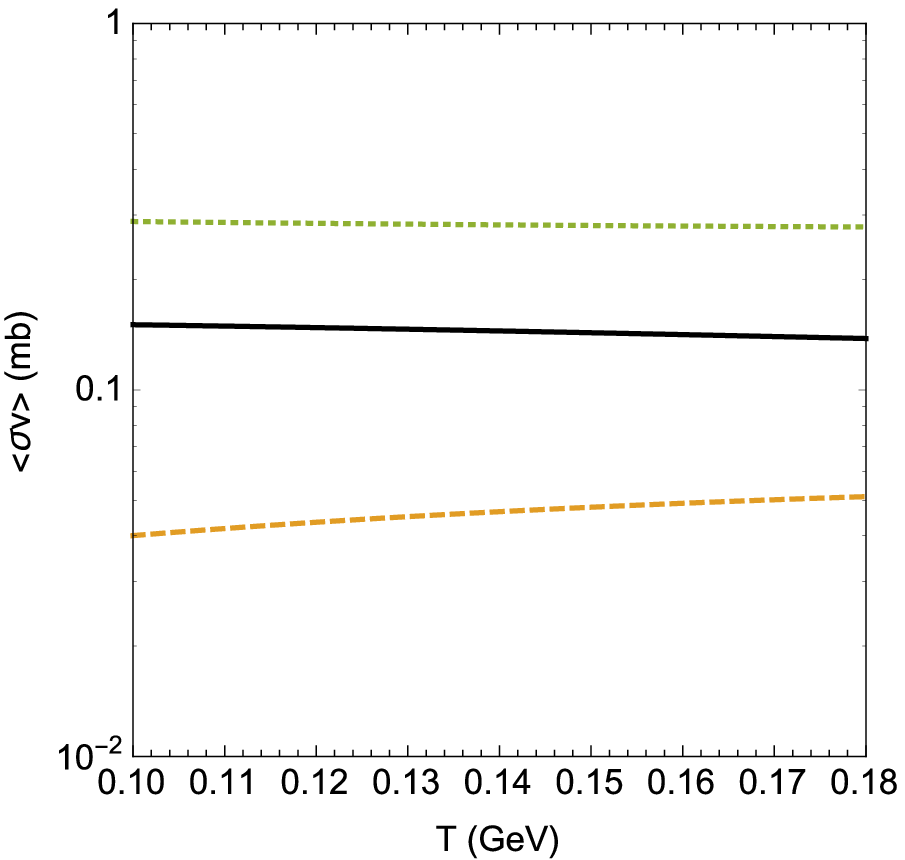}  
\caption{Thermally averaged cross sections for $\rho \Upsilon $ 
absorption and production as a function  of the temperature.  
Top panel:   $\rho \Upsilon$  in the initial state.   
Bottom panel:  $\rho \Upsilon$  in the final state. 
Solid, dashed and dotted lines represent the reactions  
with $ \bar{B} B $,  $  \bar{B}^{\ast} B $ and  
$  \bar{B}^{\ast} B^{\ast} $, respectively, in final or initial state.}
\label{thermalrho}  
\end{figure}

In Fig.~\ref{thermalrho} we plot the thermally 
averaged cross sections for the $\rho \Upsilon $ absorption and production. 
As before, in general the production reactions have larger cross 
sections than the corresponding inverse reactions.  The absorption and  
production reactions which start or end with $ \bar{B} ^{\ast} B$ and     
$\bar{B}^{\ast} B^{\ast} $ are comparable at high temperatures, while the 
production from the  $ \bar{B} B$ initial state is stronger than the 
absorption with the $ \bar{B} B$ final state.

\section{Time evolution of the $J/\psi$ abundance }
\label{TimeEv}

The present study will be completed by addressing the time evolution of the 
$\Upsilon$ abundance in hadronic matter, 
using the thermally averaged cross sections estimated in the 
previous  section.  We shall make use of the evolution equation for 
the  abundances 
of particles included in processes  discussed above. The momentum-integrated 
evolution equation has the form~
\cite{EXHIC,Cho:2017dcy,Cho:2010db,Cho:2015qca,nosx1,ChenPRC,ChoLee1}: 

\ben
\frac{d N_{\Upsilon} (\tau)}{d \tau} & = &  \sum_{\varphi = \pi, \rho  } 
\left[ \langle \sigma_{   \bar{B}  B \rightarrow \varphi \Upsilon } v_{ \bar{B}  
B } \rangle n_{\bar{B}} (\tau) N_{B} (\tau) 
 \right. \nonumber \\ 
 & & \left. 
 - \langle \sigma_{\varphi \Upsilon \rightarrow  \bar{B}  B } 
v_{ \varphi \Upsilon} 
\rangle n_{\varphi} (\tau) N_{\Upsilon}(\tau)  
\right. \nonumber \\ 
& &  \left. 
+ \langle \sigma_{  \bar{B} ^{\ast} B ^{\ast}  \rightarrow 
\varphi \Upsilon } v_{ \bar{B} ^{\ast} B ^{\ast}} 
\rangle n_{\bar{B} ^{\ast}} (\tau) N_{B ^{\ast}}(\tau) 
\right. \nonumber \\ 
& & \left.
- \langle \sigma_{\varphi \Upsilon \rightarrow  \bar{B} ^{\ast}      
B ^{\ast}  } v_{ \varphi \Upsilon} \rangle n_{\varphi} (\tau)  
N_{\Upsilon}(\tau) \right. \nonumber \\  & &  \left. + \langle 
\sigma_{  \bar{B} ^{\ast} B   \rightarrow \varphi \Upsilon } 
v_{ \bar{B} ^{\ast} B } \rangle n_{\bar{B} ^{\ast}} (\tau) 
N_{B }(\tau)  
\right. \nonumber \\ 
& & \left. 
- \langle \sigma_{\varphi \Upsilon \rightarrow           
\bar{B} ^{\ast} B  } v_{ \varphi \Upsilon} \rangle n_{\varphi} (\tau) 
N_{\Upsilon}(\tau) 
\right. \nonumber \\ 
& &  \left. 
+ \langle \sigma_{   
\bar{B}  B ^{\ast}  \rightarrow \varphi \Upsilon } v_{ \bar{B}  
B ^{\ast} } \rangle n_{\bar{B} } (\tau) N_{B ^{\ast}}(\tau)  
\right. \nonumber \\ 
& & \left.
- \langle \sigma_{\varphi \Upsilon \rightarrow  \bar{B}  B ^{\ast}  } 
v_{ \varphi \Upsilon} \rangle n_{\varphi} (\tau) N_{\Upsilon}(\tau)   \right] 
,
\label{rateeq}
\een
where $n_{i} (\tau)$ are $N_{i}(\tau)$ denote  the density and          
the abundances of  $ \pi,\rho$ and bottom mesons in hadronic matter at  
proper time $\tau$. From 
Eq.~(\ref{rateeq}) we notice that 
the $ \Upsilon $ abundance at a proper time $\tau$ depends on 
the $\varphi \Upsilon$ 
dissociation rate as well as on the $\varphi \Upsilon$  production rate.  
 We will assume that $\pi, \rho, B$ and  
$B^{\ast}$ are in equilibrium. Therefore the density $n_{i} (\tau)$ can be 
written  
as~\cite{EXHIC,Cho:2017dcy,Cho:2010db,Cho:2015qca,nosx1}
\ben n_{i} (\tau) &  \approx & \frac{1}{2 \pi^2}\gamma_{i} g_{i} m_{i}^2 
T(\tau)K_{2}\left(\frac{m_{i} }{T(\tau)}\right), 
\label{densities}
\een
where $\gamma _i$ and $g_i$ are the fugacity factor and the  
degeneracy factor of the relevant particle, respectively. The multiplicity 
$N_i (\tau)$ is 
obtained by multiplying the density $n_i(\tau)$ by the volume $V(\tau)$. 
The time dependence is introduced through the temperature, $T(\tau)$, and 
volume, $V(\tau)$, profiles, appropriate to model the dynamics of relativistic 
heavy ion collisions after the end of the quark-gluon plasma phase. The 
hydrodynamical expansion and cooling of the hadron gas are described as  in 
Refs.~\cite{EXHIC,Cho:2017dcy,Cho:2010db,Cho:2015qca,nosx1,ChenPRC,ChoLee1}, 
which are based on the boost invariant Bjorken picture 
with an accelerated transverse expansion:
\ben
T(\tau) & = & T_C - \left( T_H - T_F \right) \left( \frac{\tau - \tau _H }
{\tau _F -  
\tau _H}\right)^{\frac{4}{5}} , \nonumber \\V(\tau) & = & \pi \left[ R_C + v_C 
\left(\tau - \tau_C\right) + \frac{a_C}{2} \left(\tau - \tau_C\right)^2 
\right]^2 \tau \, c , \nonumber \\
\label{TempVol}
\een
where $R_C $ and $\tau_C$  denote the final 
transverse  and longitudinal sizes of the quark-gluon plasma; $v_C $ 
and  $a_C $ are its transverse flow velocity and transverse 
acceleration at $\tau_C $; $T_C = 175$ MeV is the critical temperature for the 
quark-gluon plasma to hadronic matter transition; $T_H = T_C = 175 $ MeV  is 
the temperature of the hadronic matter at the end of the mixed phase, occurring 
at the time $\tau_H $; and the freeze-out temperature,  $T_F = 125$ MeV,     
leads to a freeze-out time $\tau _F $. 

In the present approach  we assume that the total 
number of bottom quarks in bottom hadrons is  conserved during the 
processes. This   
number can be calculated with perturbative QCD and  yields the bottom quark 
fugacity factor $\gamma _b $ in Eq.~(\ref{densities}) 
\cite{EXHIC,Cho:2017dcy,Cho:2010db,Cho:2015qca,nosx1}. 
The total number of  pions and $\rho$ mesons at freeze-out was taken from 
Refs.~\cite{ChenPRC,nosx1,ChoLee1}.

The evolution of $\Upsilon$ multiplicity is analyzed in two scenarios: 
with the hadron gas formed in central
 $Au - Au$ collisions at $\sqrt{s_{NN}}= 200$ GeV 
at RHIC and central $Pb-Pb$ collisions at $\sqrt{s_{NN}} = 5$ TeV at the LHC. 
The parameters which we need as input in Eqs. (\ref{TempVol}) are listed 
in Ref.~\cite{EXHIC}, and are reproduced in 
Table~\ref{param} for convenience. 
Notice that the estimate of the $\Upsilon$ yield at the end of the mixed 
phase, given in the last column of  
Table~\ref{param}, is done in the context of the statistical model, in which  
hadrons are in thermal and chemical equilibrium when they are produced.
Therefore, at RHIC the  $\Upsilon$ multiplicity at $\tau _H$ is
\ben 
N_{\Upsilon} & \approx &  \frac{1}{2 \pi^2} \gamma_{b} ^2  
g_{\Upsilon} m_{\Upsilon}^2 T_H K_{2} \left(\frac{m_{\Upsilon} }{T_H}\right) 
V(\tau_H) \nonumber \\ 
&\approx & 1.705 \times 10^6 . 
\label{NJPsi}
\een 
A similar calculation for the case of LHC gives $N_{\Upsilon} \approx 0.00106 $. 

\begin{widetext}   
\begin{center}
\begin{table}[h!]
\caption{Parameters used in the parametrization of the hydrodynamical expansion,
given by Eqs.~(\ref{TempVol}). }
\vskip1.5mm
\label{param}
\begin{tabular}{c| c c c c c c c c c}
\hline
\hline
& $\sqrt{s_{NN}}$ (TeV) & $v_C$ (c) & $a_C$ (c$^2$/fm) & $R_C$ (fm) &     
$\tau_C$ (fm/c) & $\tau_H$ (fm/c)  & $\tau_F$ (fm/c) & $\gamma_b$  
& $N_{\Upsilon}$ \\   
\hline
RHIC & 0.2 & 0.4 & 0.02 & 8  &  5 & 7.5 & 17.3 & $2.2 \times 10^6$ & $1.705 
\times 10^6$ 
\\  
LHC  & 5   & 0.6 & 0.044& 13.11  & 5  & 7.5 & 20.7 & $3.3 \times 10^7$ & 0.00106 
\\  \hline
 \hline
\end{tabular}
\end{table}
\end{center}
\end{widetext}

\begin{figure}[th]
\centering
\includegraphics[width=8.0cm]{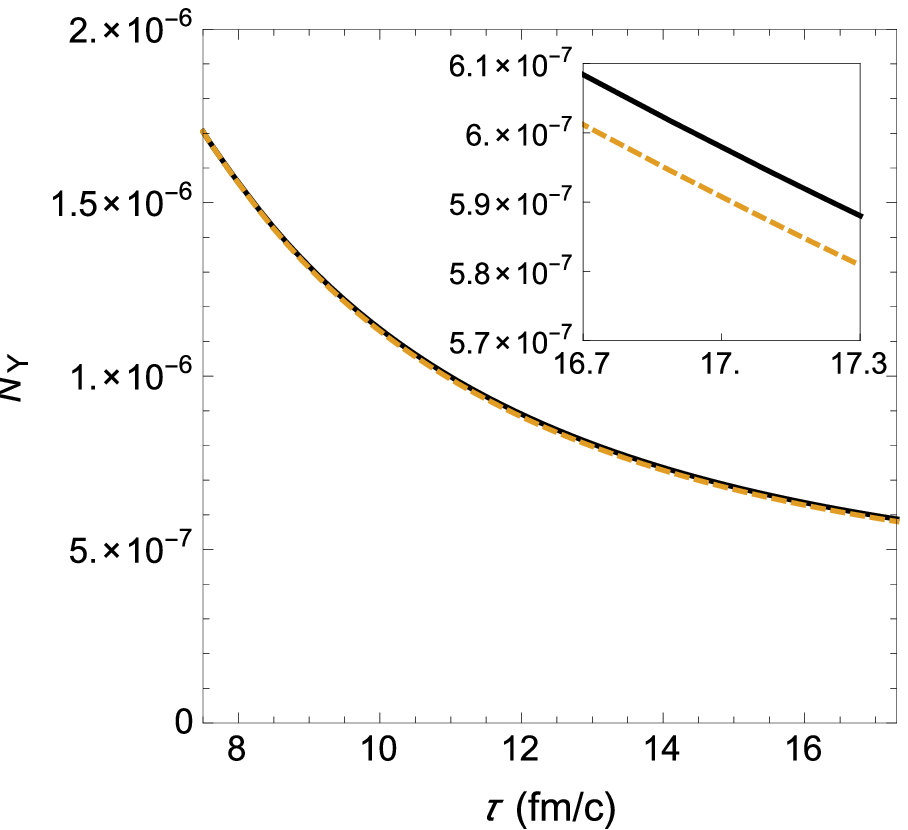}  
\\
\includegraphics[width=8.0cm]{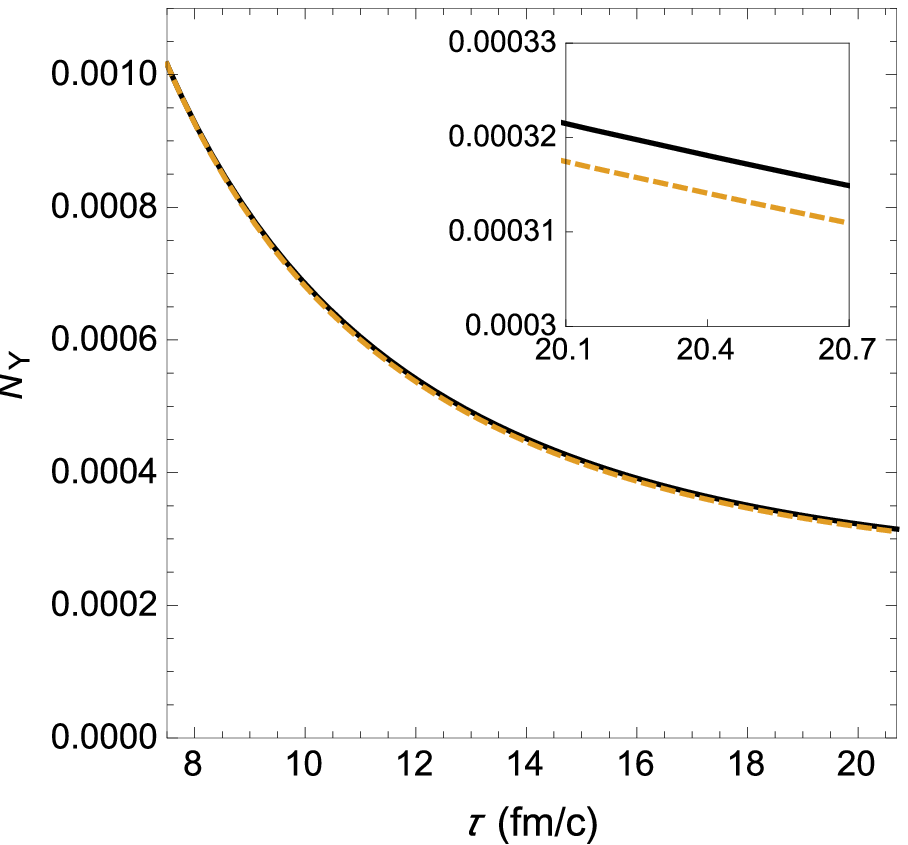}
\caption{Top: Time evolution of $\Upsilon$ abundance as a function of the proper  
time in central Au-Au collisions at $\sqrt{s_{NN}} = 200$ GeV. Solid and 
dashed lines represent the situations with only $ \pi - \Upsilon$ 
interactions and also adding the  $\rho - \Upsilon$, respectively.
Bottom: the same as on the top for LHC conditions. 
}
\label{TimeEvolUpsilon}
\end{figure}

The time evolution of the $\Upsilon$ 
abundance is plotted in Fig.~\ref{TimeEvolUpsilon} as a function of  
the proper time, for the two types of collisions discussed above: at 
RHIC (on the upper panel) and at the LHC (on the lower panel). 

The behavior of the $\Upsilon$ multiplicity observed  in 
Fig.~\ref{TimeEvolUpsilon} is not difficult to understand. Due to the assumption 
that the hadronic stage at LHC is longer compared to that at RHIC,
more bottomonium states are lost in the hadronic medium at LHC.      
Also, it can be noticed, from Eq.~(\ref{rateeq}), that the evolution    
of the $\Upsilon$ multiplicity depends on the production and absorption 
cross sections and also on the abundances  of the other mesons. Although 
the production cross sections are greater than the absorption ones, which would  
enhance the $\Upsilon$ yield, the relative meson multiplicities  lead to its 
reduction, since there are much more light mesons (especially pions) in 
the hadron gas to collide and destroy the bottomonium states than 
$B^{(\ast)}$'s and  $\bar{B}^{(\ast)}$'s to interact and create them. 
Besides, from the solid and dotted lines in Fig.~\ref{TimeEvolUpsilon} we  
can infer that the role of  the $\rho$ mesons in the gas is not relevant   
when compared to that of the pions. This comes from a cancellation between  
the terms associated to the production and absorption reactions: the 
different magnitudes of production and absorption processes are compensated 
by the relative multiplicities.

The results shown in Fig.~\ref{TimeEvolUpsilon} suggest a decrease of the 
$\Upsilon$ yield of almost $\simeq 66$ \%  at RHIC and 
$\simeq 70$ \% at the LHC.  These numbers   
are compatible with (\ref{raaups}). Taken literally, they  would suggest that 
all the suppression comes from the hadron gas phase.  However we are not yet 
in the position of sustaining this strong statement. Before that, there is a 
number of points to be discussed. First the interactions in the reactions are 
naturally dependent  on the effective formalism considered, which determines  
the magnitudes of the cross sections. A change  in the magnitude 
of  the production reactions will modify those of  the absorption in the same 
proportion. This will lead to an overall multiplicative factor in the right 
hand side of rate equation, Eq.~(\ref{rateeq}), modifying the curves in 
Fig.~\ref{TimeEvolUpsilon}. Besides, our results are strongly dependent on 
the form factors and cutoff values: different choices would  modify the slope 
of  the curves in Fig.~\ref{TimeEvolUpsilon}. Furthermore,  the relevance  
of the parametrization of the hydrodynamical expansion exhibited in 
Eq. (\ref{TempVol})  can not  be underestimated.  
Different parameters can make the system cool faster or slower and 
accordingly change the multiplicities of the distinct particles. 

\begin{figure}[th]
\centering
\includegraphics[width=8.0cm]{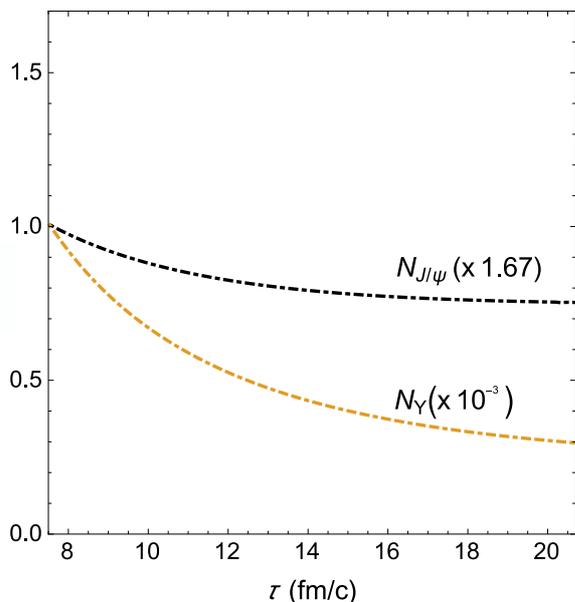}
\caption{Top: Time evolution of   $J/\psi$ (upper line) and 
$\Upsilon$ (lower line) abundances as a function of the proper  
time in central Pb-Pb collisions at  the LHC. }
\label{TimeEvolComp}
\end{figure}

Notwithstanding the  points raised above, we stress the main result of 
this work: a reduction of  the number of $\Upsilon$'s in the hadron gas,  
which seems to be larger than in the case of $J/\psi $ reported in  
Ref.~\cite{akhematonani}. Before closing this section, we show in 
Fig.~\ref{TimeEvolComp}  a comparison between the $\Upsilon$ and 
$J/\psi$ multiplicities as a function 
of the proper time. The latter one was already published in 
\cite{akhematonani}.  For the sake of comparison we have rescaled them to the 
unity at the initial time. The $J/\psi$ suppression is only of 
$\simeq 25$ \%, whereas it is of $\simeq 70$ \% in the case of the 
$\Upsilon$.

\section{Concluding Remarks}
\label{Conclusions}

In this work  we have  analyzed 
the hadronic effects on the $\Upsilon $ abundance in heavy ion collisions. 
Effective Lagrangians 
have been used to calculate  
the cross sections for the $\Upsilon$-production processes  
$\bar{B}^{(*)} + B^{(*)}  \to \Upsilon + (\pi, \rho)$,  and 
also for the corresponding inverse processes associated to the 
$\Upsilon$ absorption.  
We have  also computed the thermally averaged cross sections for 
the dissociation and production 
reactions. Finally, 
we have employed the thermally averaged cross sections as inputs in  
the rate equation and have 
determined the time 
evolution of the $\Upsilon$ abundance in a hadron gas.

Examining the existing literature on cross section calculations, the 
present work has 
introduced the following improvements: 
inclusion of reactions which start or end with $ \bar{B} B$ and 
$\bar{B}^{\ast} B^{\ast} $ in the 
case of the pion-$\Upsilon$ scattering,  and  $\bar{B}^{\ast} B $ in 
that involving $\rho$ meson;  inclusion of 
the anomalous parity interactions  processes in the effective 
Lagrangian approach.

Our results suggest that the interactions between $\Upsilon$ and light mesons 
reduce the  $\Upsilon$ abundance at the end of 
the quark gluon plasma phase by $\simeq 70$ \%,  which is more 
than in the case of the $J / \psi $ reported in Ref.~\cite{akhematonani}.

In conclusion, despite the fact that there 
are points to be improved to obtain a more realist description 
of the HIC phenomenology, we believe that our findings are    
important for the physics of both the  quark gluon plasma and 
hadronic phases.  Our result should encourage further studies of
the $\Upsilon$ suppression in the hadron gas phase of relativistic heavy
ion collisions.

\begin{acknowledgements}

The authors would like to thank the Brazilian funding agencies CNPq (contracts
310759/2016-1, 311524/2016-8, 308088/2017-4 and 400546/2016-7),   
FAPESP (contract  17/07278-5) 
FAPESB (contract INT0007/2016) for financial support.

\end{acknowledgements} 

\begin{thebibliography}{99}



\bibitem{qgpdisc}      J.~Adams {\it et al.} [STAR Collaboration],
                       Nucl.\ Phys.\ A {\bf 757}, 102 (2005).

\bibitem{rev-qgp}      For a recent review, see   
                       P. Braun-Munzinger, V. Koch,
                       T. Schafer, and J. Stachel,
                       Phys. Rept. {\bf 621}, 76 (2016).

\bibitem{matsatz}      T.~Matsui and H.~Satz,
                       Phys.\ Lett.\ B {\bf 178}, 416 (1986).

\bibitem{sup-rev-rapp}     For a review , see  R. Rapp,
                           Prog.  Part.  Nucl. Phys. {\bf 65}, 209 (2010).

\bibitem{hf-rev}       A.~Andronic {\it et al.},
                       Eur.\ Phys.\ J.\ C {\bf 76}, 107 (2016) and 
                       references  therein.

\bibitem{regraf}       R. L. Thews, M. Schroedter, and J. Rafelski, 
                       Phys. Rev. C {\bf 63},   054905 (2001); 
                       P. Braun-Munzinger and J. Stachel, 
                       Phys. Lett. B {\bf 490}, 196 (2000).

 
\bibitem{Abelev:2013ila}    B.~B. Abelev  \emph{et~al.} (Collaboration ALICE), 
                            Phys. Lett. B {\bf 734}, 314 (2014).

\bibitem{Adam:2016rdg}      J. Adam \emph{et~al.} (Collaboration ALICE), 
                            Phys. Lett. B {\bf 766}, 212 (2017).

\bibitem{Zha:2017xsm}       W. Zha and Z. Tang, 
                            Nucl. Part. Phys. Proc. {\bf 289}, 83 (2017).

\bibitem{durapphe}          X.~Du, R.~Rapp and M.~He,
                            Phys.\ Rev.\ C {\bf 96}, 054901 (2017). 

\bibitem{regebot}         A. Andronic, P. Braun-Munzinger, 
                          K. Redlich, and J. Stachel, 
                          Phys. Lett. B {\bf 652}, 259 (2007); 
                          M. I. Gorenstein, A. P. Kostyuk, 
                          H. Stoecker, and W. Greiner, 
                          Phys. Lett. B {\bf 509}, 277 (2001). 

\bibitem{chineses}        Y. Liu et al., Phys. Lett. B {\bf 697}, 32 (2011); 
                          K. Zhou, N. Xu and P. Zhuang, 
                          Nucl. Phys. A {\bf 931}, 654 (2014).

\bibitem{emerapp}         A.~Emerick, X.~Zhao and R.~Rapp,
                          Eur.\ Phys.\ J.\ A {\bf 48}, 72 (2012).

\bibitem{mike1}           B. Krouppa and M. Strickland, 
                          Universe { \bf 2}, 16 (2016). 

\bibitem{mike2}           B. Krouppa, R. Ryblewski, and M. Strickland, 
                          Phys. Rev. C {\bf 92}, 061901 (2015).

\bibitem{leonardo}        Z.~Hu, N.~T.~Leonardo, T.~Liu and M.~Haytmyradov,
                          Int.\ J.\ Mod.\ Phys.\ A {\bf 32}, 1730015 (2017).

\bibitem{cms}             CMS Collab. (S. Chatrchyan et al.), 
                          Phys. Rev. Lett. {\bf 107}, 052302 (2011); 
                          Z. Hu, J. Phys. G  {\bf 38}, 124071 (2011);
                          CMS Collab. (S. Chatrchyan et al.), 
                          J. High Energy Phys. {\bf 05}, 063 (2012).


\bibitem{upsup-star}      L.~Adamczyk {\it et al.} [STAR Collaboration],
                          Phys.\ Lett.\ B {\bf 735}, 127 (2014);
                          Erratum: [Phys.\ Lett.\ B {\bf 743}, 537 (2015)].

\bibitem{cms-psi}         V.~Khachatryan {\it et al.} [CMS Collaboration],
                          Eur.\ Phys.\ J.\ C {\bf 77}, 252 (2017).




\bibitem{cms-ups-18}      A.~M.~Sirunyan {\it et al.} [CMS Collaboration],
                          arXiv:1805.09215 [hep-ex].

\bibitem{cms-ups-17}      V.~Khachatryan {\it et al.} [CMS Collaboration],
                          Phys.\ Lett.\ B {\bf 770}, 357 (2017).

\bibitem{cms-ups-17a}     V. Khachatryan {\it  et al.} [CMS Collaboration],
                          Phys.\ Lett.\ B {\bf 04}, 031 (2017).

\bibitem{hg-pbm}          V.~M.~Shapoval, P.~Braun-Munzinger and Y.~M.~Sinyukov,
                          Nucl.\ Phys.\ A {\bf 968}, 391 (2017). 

\bibitem{hg-alice}        J.~Adam {\it et al.} [ALICE Collaboration], 
                          production at  in Pb-Pb collisions 
                          Phys.\ Rev.\ C {\bf 95}, 064606 (2017); 
                          B.~B.~Abelev {\it et al.} [ALICE Collaboration],
                          Phys.\ Rev.\ C {\bf 91}, 024609 (2015).


\bibitem{hg-kst}          B. B. Abelev et al. (The ALICE Collaboration), 
                          Phys. Rev. C {\bf 91}, 024609 (2015). 


\bibitem{hg-rho}          V. G. Riabov et al. (The ALICE Collaboration), 
                          J. Phys. Conf. Ser. {\bf 798}, 012054 (2017); 
                          C. Markert et al. (The ALICE Collaboration),
                          J. Phys. Conf. Ser. {\bf 878}, 012003 (2017).
         

\bibitem{akhematonani}     L. M. Abreu, K. P. Khemchandani, A. Mart\'{\i}nez
                           Torres, F. S. Navarra and M. Nielsen,
                           Phys. Rev. C {\bf 97}, 044902 (2018).

 
\bibitem{psipi-ft1}        S. G. Matinyan and B. M\"{u}ller,
                           Phys. Rev. C {\bf 58}, 2994 (1998); 
                           A.~Bourque and C.~Gale,  
                           Phys.\ Rev.\ C {\bf 80}, 015204 (2009);  
                           Phys.\ Rev.\ C {\bf 78}, 035206 (2008);
                           A.~Bourque, C.~Gale and K.~L.~Haglin,  
                           Phys.\ Rev.\ C {\bf 70}, 055203 (2004);
                           Z.~Lin and C.~M.~Ko,   
                           Phys.\ Rev.\ C {\bf 62}, 034903 (2000);
                           J. Phys. G {\bf 27}, 617 (2001);   
                           F. S. Navarra, M. Nielsen and M. R. Robilotta, 
                           Phys. Rev. C {\bf 64}, 021901(R) (2001);
                           K. L. Haglin and C. Gale, 
                           Phys. Rev. C {\bf 63}, 065201 (2001); 
                           K. L. Haglin, Phys. Rev. C {\bf 61}, 031902 (2000);
                           F.~Carvalho, F.~O.~Duraes, F.~S.~Navarra and 
                           M.~Nielsen, Phys.\ Rev.\ C {\bf 72}, 024902 (2005). 

\bibitem{psipi-oh}         Y. Oh, T. Song and S. H. Lee,
                           Phys. Rev. C {\bf 63}, 034901 (2001).



\bibitem{kmnn}             K.~P.~Khemchandani, A.~Martinez Torres, 
                           M.~Nielsen and F.~S.~Navarra, 
                           Phys.\ Rev.\ D {\bf 89}, 014029 (2014).

 

\bibitem{psipi-sr}         F.~O.~Duraes, H.~c.~Kim, S.~H.~Lee, F.~S.~Navarra
                           and M.~Nielsen,
                           Phys.\ Rev.\ C {\bf 68}, 035208 (2003);
                           F.~O.~Duraes, S.~H.~Lee, F.~S.~Navarra
                           and M.~Nielsen,
                           Phys.\ Lett.\ B {\bf 564}, 97 (2003);
                           F.~S.~Navarra, M.~Nielsen, R.~S.~Marques
                           de Carvalho and G.~Krein,
                           Phys.\ Lett.\ B {\bf 529}, 87 (2002).
                           
\bibitem{Upsilon1}         Z. Lin and C. M. Ko, 
                           Phys. Lett. B {\bf 503}, 104 (2001). 
                 
\bibitem{rocaoset}         H. Nagahiro, L. Roca and E. Oset,
                           Eur. Phys. J. A {\bf 36}, 73 (2008).
            

\bibitem{nosx1}            A.~Martinez Torres, K.~P.~Khemchandani,
                           F.~S.~Navarra, M.~Nielsen and L.~M.~Abreu,
                           Phys.\ Rev.\ D {\bf 90}, 114023 (2014)
                           Erratum: [Phys.\ Rev.\ D {\bf 93}, 059902 (2016)].

\bibitem{nosx2}            L. M. Abreu, K. P. Khemchandani, A. Martinez Torres,
                           F. S. Navarra and M. Nielsen,
                           Phys. Lett.  B {\bf 761}, 303 (2016).


\bibitem{Pearce:1990uj}    B.~C.~Pearce and B.~K.~Jennings,
                           Nucl.\ Phys.\ A {\bf 528}, 655 (1991).


\bibitem{Oh:2002vg}        Y.~s.~Oh, T.~s.~Song, S.~H.~Lee and C.~Y.~Wong,
                           J.\ Korean Phys.\ Soc.\  {\bf 43}, 1003 (2003). 

\bibitem{Ronchen:2012eg}   D.~Ronchen {\it et al.},
                           Eur.\ Phys.\ J.\ A {\bf 49}, 44 (2013).
  
\bibitem{Koch}             P. Koch, B. Muller and J. Rafelski, 
                           Phys. Rep. {\bf 142}, 167 (1986).

\bibitem{Cho:2010db}       S.~Cho {\it et al.} (ExHIC Collaboration), 
                           Phys.\ Rev.\ Lett.\  {\bf 106}, 212001 (2011). 

\bibitem{Cho:2015qca}      S. Cho and S. H. Lee, 
                           Phys.\ Rev.\ C {\bf 97}, 034908 (2018).

\bibitem{EXHIC}            S. Cho {\it et al.} (ExHIC Collaboration), 
                           Phys. Rev. C {\bf 84}, 064910 (2011).

\bibitem{Cho:2017dcy}      S.~Cho {\it et al.} (ExHIC Collaboration),
                           Prog.\ Part.\ Nucl.\ Phys.\  {\bf 95}, 279 (2017).

\bibitem{ChenPRC}          L. W. Chen, C. M. Ko, W. Liu and M. Nielsen,  
                           Phys. Rev. C {\bf 76}, 014906 (2007).

\bibitem{ChoLee1}          S.~Cho and S.~H.~Lee, 
                           Phys. Rev. C {\bf 88}, 054901 (2013) . 

\end{thebibliography}
%
%

\end{document}